\documentclass[%
 aip,
 amsmath,amssymb,
 reprint,%
]{revtex4-1}
\usepackage{graphicx}
\usepackage{dcolumn}
\usepackage{bm}
\usepackage{xcolor}
\usepackage{comment}

\usepackage[utf8]{inputenc}
\usepackage[T1]{fontenc}
\usepackage{mathptmx}
\usepackage{etoolbox}
\makeatletter
\def\@email#1#2{%
 \endgroup
 \patchcmd{\titleblock@produce}
  {\frontmatter@RRAPformat}
  {\frontmatter@RRAPformat{\produce@RRAP{*#1\href{mailto:#2}{#2}}}\frontmatter@RRAPformat}
  {}{}
}%
\makeatother

\begin{document}

\preprint{AIP/123-QED}

\title[Stochastic control of spiking activity bump expansion: monotonic and resonant phenomena]{Stochastic control of spiking activity bump expansion: monotonic and resonant phenomena}
\author{Anna Zakharova}
\affiliation{Institut f\"{u}r Theoretische Physik, Technische Universit\"{a}t Berlin, 10623 Berlin, Germany}
\affiliation{Bernstein Center for Computational Neuroscience, Humboldt-Universit\"{a}t zu Berlin, Philippstra{\ss}e 13, 10115 Berlin, Germany}

\author{Vladimir V. Semenov}
\email{semenov.v.v.ssu@gmail.com}
\affiliation{Institut f\"{u}r Theoretische Physik, Technische Universit\"{a}t Berlin, 10623 Berlin, Germany}
\affiliation{Institute of Physics, Saratov State University, 410012 Saratov, Russia}

\date{\today}

\begin{abstract}

We consider spatially localized spiking activity patterns, so-called bumps, in ensembles of bistable spiking oscillators. The bistability consists in the coexistence of self-sustained spiking dynamics and quiescent steady-state regime. We show numerically that the processes of growth or contraction of such patterns can be controlled by varying the intensity of multiplicative noise. In particular, the effect of the noise is monotonic in an ensemble of the coupled Hindmarsh-Rose oscillators. On the other hand, in another model proposed by V. Semenov et al. in 2016 (see Ref. \cite{semenov2016}), a resonant noise effect is observed. In that model, stabilization of the activity bump expansion is achieved at an appropriate noise level, and the noise effect reverses with a further increase in noise intensity. Moreover, we show the constructive role of nonlocal coupling which allows to save domains and fronts being totally destroyed due to the action of noise in the case of local coupling. 

\end{abstract}

\maketitle

\section{Introduction}
\label{intro}
Formation of self-sustained states localized in space, so-called localised activity bumps, is a frequent phenomenon in biological neural networks \cite{coombes2014}. In particular, such states have been reported by neurophysiologists and associated with such brain functions as working memory \cite{compte2000,wimmer2014} and head-direction control \cite{taube1998,sharp2001}. A broad spectrum of mathematical models is used for the investigation of such structures in different forms: stationary soliton solution \cite{blomquist2005,yousaf2013}, localised oscillating structures \cite{rubin2004,rubin2005,rubin2006,laing2015,avitabile2017,schmidt2020}, propagating and expanding patterns \cite{avitabile2015}. In the present work, the existence and expansion of spiking activity bumps in ensembles of bistable oscillators are explored. Here, the property of bistability in the studied models results from the coexistence of a stable limit cycle and a stable steady state in the phase space of an individual (uncoupled) oscillator.

It is well known that bistable spatially-extended systems can exhibit dynamics in which two domains with propagating fronts between them are formed in space. Such fronts occur frequently in chemistry, for instance, in the Schl{\"o}gl model \cite{schloegl1972,schloegl1983,loeber2014} developed to explain an autocatalytic reaction mechanism, as well as in various other fields, such as electronics \cite{schoell2001} or flame propagation theory \cite{zeldovich1938}, to name just a few. Similar processes associated with domain expansion and called 'coarsening' are studied in the context of physics of liquid crystals \cite{yurke1992}, magnetism \cite{bray1994,cugliandolo2010,caccioli2008,denholm2019}, physics and chemistry of materials \cite{goh2002,zhang2019,zhang2019-2,geslin2019}, laser physics \cite{yanchuk2012,marino2014,javaloyes2015}, electronics \cite{semenov2018}, and animal population statistics \cite{dobramysl2018}. It should be noted that besides bistable spatially-extended systems \cite{bray1994}, such effects have been observed for time-delay models \cite{yanchuk2012,marino2014,semenov2018,ruschel2019}.

Since the phenomena referred to as wavefront propagation and coarsening are observed in a wide spectrum of dynamical systems, the development of principles to control these effects represents a relevant interdisciplinary problem. One of the known approaches involves the impact of noise for this purpose. Noise is actively used for the control of various phenomena such as, for example, synchronization in multilayer networks where the so-called multiplexing noise included into the interaction of the network layers regulates the synchronization of spatio-temporal patterns \cite{VAD20}. Moreover, constructive role of noise and the related resonance phenomena continue to attract the attention of researches as confirmed by recent works on, for example, coherence resonance \cite{MAS17,bogatenko2018,SEM18,PIS19,YAM19,MAS21} or stochastic resonance \cite{SEM22,YAM21}. 

In the context of front propagation in bistable media, an effective control is achieved by applying multiplicative noise \cite{engel1985} and relies on the fact that multiplicative noise has a systematic impact on the symmetry properties of the reaction-term \cite{garcia-ojalvo1999,mendez2011}. This  multiplicative-noise-based scheme is also used in the current study. 

We show that the effects of noise on propagating phenomena take more complicated forms in bistable systems where oscillatory and steady-state dynamics coexists compared to bistable systems with two coexisting stable steady states, whereas the ability of multiplicative noise to control the propagation speed and direction of the front remains. 
In particular, we demonstrate numerically that the noise effect can be either monotonic or resonant, depending on the intrinsic properties of a partial oscillator in the ensemble. For this purpose, we consider two examples. 
The first example exhibits a monotonic effect of noise, where the increase in noise intensity first slows and stops the front propagation and then reverses its direction. In contrast, the effect observed in the second model has a resonant character, where the front propagation is slowed down and stabilized due to the action of noise in a finite range of noise intensity.

\section{Monotonic impact of noise}
\begin{figure}[t]
\centering
\includegraphics[width=0.48\textwidth]{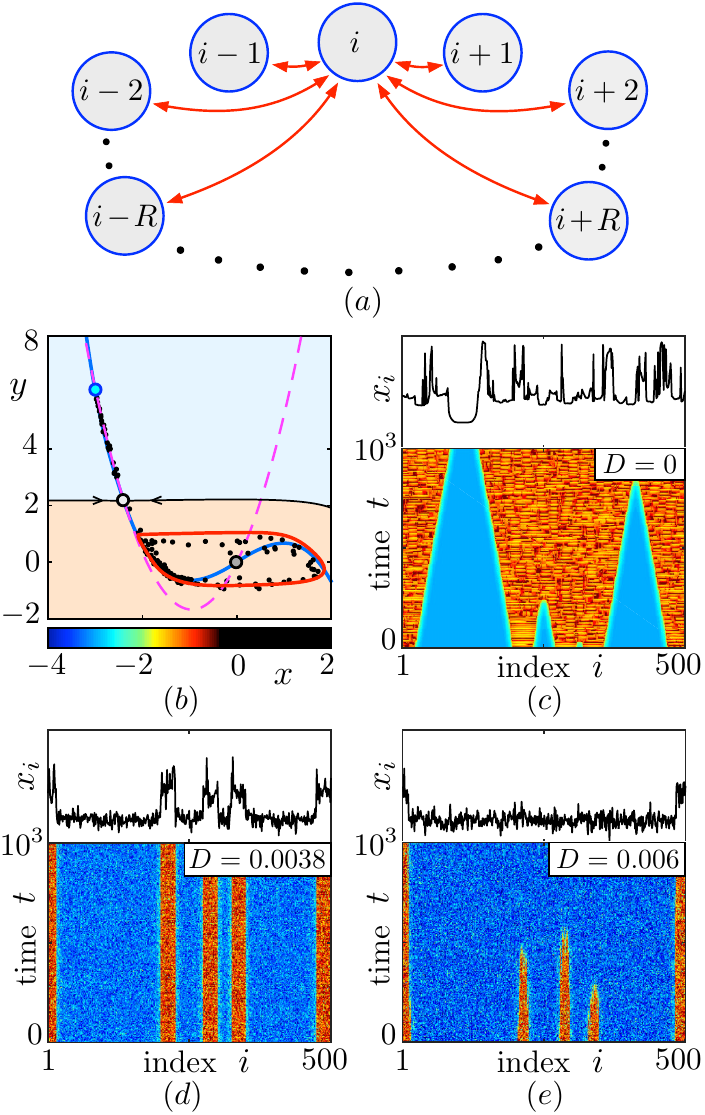} 
\caption{(a) Schematic representation of a ring of nonlocally coupled oscillators; (b)-(e) System Eqs. (\ref{HR_ensemble}): (b) Partial oscillator's phase space for $\varepsilon=0.05$, $a=1$, $b=1.9$, $c=0.55$. Two attractors coexist: a stable limit cycle (red closed curve) and a stable equilibrium point (blue circle). The translucent red and blue regions display the basin of attraction for the limit cycle and the equilibrium point, respectively. The light grey circle represents a saddle fixed point, black arrowed curves indicate the saddle separatrix. The dark grey circle marks an unstable steady state. Nullclines are shown by the blue solid ($dx/dt=0$) and magenta dashed ($dy/dt=0$) curves; (c)-(e) Space-time plots illustrating propagating phenomena for varying noise intensity for the fixed coupling strength $\sigma=0.3$ and coupling radius $R=10$. The partial oscillator parameters are the same as for panel (b). The upper insets show the ensemble state at the last moment $t=10^3$.}
\label{fig1}
\end{figure}

First, we consider a ring of the nonlocally coupled Hindmarsh-Rose oscillators (schematically illustrated in Fig.~\ref{fig1}(a)) with multiplicative noise. A two-dimensional modification of the Hindmarsh-Rose oscillator described in Ref. \cite{tsuji2007} was chosen as a partial oscillator:
\begin{equation}
\label{HR_ensemble}
\begin{array}{rl}
\varepsilon \dfrac{dx_{i}}{dt}= & \left(a+\sqrt{2D}n_{i}(t) \right)x_{i}-\dfrac{x^{3}_{i}}{3}-y_{i}\\
 & +\dfrac{\sigma}{2R}\sum\limits^{i+R}_{j=i-R}(x_j-x_i),\\
\dfrac{dy_{i}}{dt}= &x^{2}_{i}+bx_{i}-cy_{i},
\end{array}
\end{equation}
where $\varepsilon$, $a$, $b$, $c$ are the partial oscillator's parameters, $R$ is a coupling radius, $\sigma$ is the coupling strength, $i=1,...,N$. Further, $n_{i}(t) \in \mathbb{R}$ is Gaussian white noise, i.e. $\langle n_i(t)\rangle=0$ and $\langle n_i(t)n_j(t')\rangle =\delta_{ij}\delta(t-t')$ for all $i,j$, and $D$ is the noise intensity. The ensemble is studied for periodic boundary conditions (ring), i.e., the indices are considered modulo $N$. Numerical modelling of ensemble (\ref{HR_ensemble}) and further model (\ref{SN_ensemble}) is carried out by integration of studied differential equations using the Heun method \cite{mannella2002} with the time step $\Delta t=0.0001$.

In this work we fix the parameters $\varepsilon=0.05$, $a=1$, $b=1.9$, and $c=0.55$, for which the partial oscillator represents a bistable dynamical system. The corresponding structure of the partial oscillator phase space is depicted in Fig.~\ref{fig1}(b). Two attractors coexist in the phase space: a stable limit cycle (red closed curve) and a stable steady state (blue circle), whose basins of attraction (translucent blue and red regions) are separated by the separatrix (black arrowed curves) of the saddle (light grey circle). Two nullclines ${dx}/{dt}=0$ (blue solid line) and ${dy}/{dt}=0$ (magenta dashed line) intersect such that there exists the third equilibrium point (dark grey circle) which is unstable.

The ensemble Eqs. (\ref{HR_ensemble}) of $N=500$ oscillators is studied for the bistable regime and the fixed coupling radius $R=10$. If the initial conditions are chosen such that both basins of attraction of the stable limit cycle and the equilibrium point are involved, then two kinds of domains corresponding to the self-oscillatory spiking activity and the quiescent steady state regime are formed. After that, fronts separating the domains begin to propagate such that the domains of spiking activity invade the entire space [Fig.~\ref{fig1}~(c)]. The instantaneous state of system Eqs. (\ref{HR_ensemble}) is depicted by black points in 
[Fig.~\ref{fig1}~(b)]. Thus, the collective spiking activity of oscillators is associated with the slow-fast motions tracing trajectories close to the stable limit cycle existing in the phase space of the single oscillator. As can be seen in Fig.~\ref{fig1}~(c), the collective behaviour in spiking activity domains has asynchronous character which is a manifestation of spatial chaos. 

\begin{figure}[t]
\centering
\includegraphics[width=0.48\textwidth]{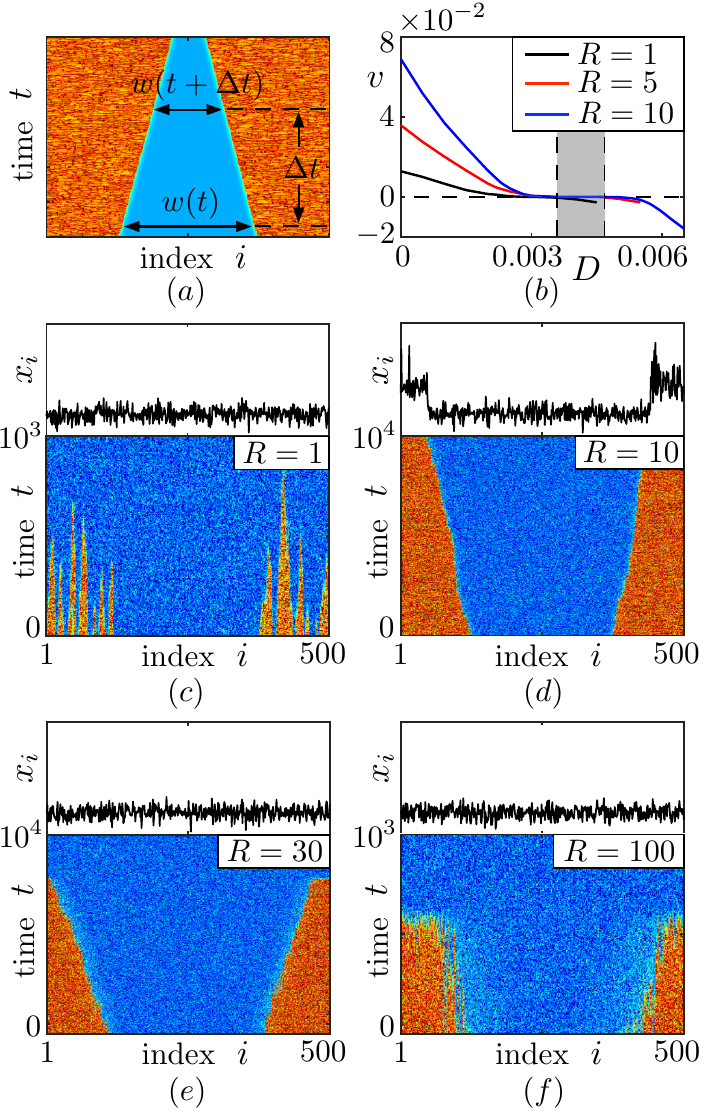} 
\caption{Dynamics in system Eqs. (\ref{HR_ensemble}): (a) Illustration for the calculation of the front propagation speed $v$ on an example of a single domain; (b) Dependence of the front propagation speed on the noise intensity $v(D)$ obtained for different values of the coupling range $R$; (c)-(f) Spatio-temporal dynamics for the fixed value of noise intensity $D=0.006$, the same initial conditions and different values of the coupling radius $R$.  The space-time plots in panels (a) and (c)-(f) are obtained in numerical simulations starting from the same initial conditions. The upper insets show the ensemble state at the last moments $t=10^3$ or $t=10^4$. Other parameters: $\varepsilon=0.05$, $a=1$, $b=1.9$, $c=0.55$, $\sigma=0.3$. }
\label{fig2}
\end{figure}

In the presence of multiplicative noise the front propagation speed  depends on the noise intensity. As illustrated in Fig.~\ref{fig1}~(c)-(e), it slows down with the noise intensity increasing, so that the activity bump expansion stops for a certain noise level, see  Fig.~\ref{fig1}~(d). 
Further increasing of the noise strength inverts the front propagation and the quiescent steady state regime invades the whole space Fig.~\ref{fig1}~(e). To visualise the impact of noise, we introduce the front propagation speed as
$v=(w(t) - w(t+\Delta t))/2\Delta t$. Here, $w(t)$ and $w(t+\Delta t)$ are the activity bump widths at the moments $t$ and  $t+\Delta t$, respectively (illustrated in Fig.~\ref{fig2}~(a)). 
Figure~\ref{fig2}~(b) shows the dependence of front velocity on the noise intensity $D$ for different values of the coupling range $R$. Let us discuss the curve $v(D)$ for $R=10$. First, the dependence $v(D)$ is almost linear (in the range $D\in[0,0.002]$). After that, the dependence becomes  almost flat, such that the propagation speed tends to zero in the range $D\in[0.0035,0.0046]$ (grey region in Fig.~\ref{fig2}~(b)). The negative values of $v(D)$ mean that the front propagation is inverted.

Another interesting observation is that non-local coupling can play a constructive role in stabilizing fronts. 
This is manifested as sustaining the propagating fronts for a higher noise level, whereas the fronts collapse in the presence of local coupling. 
This is shown in Fig.~\ref{fig2}(c)-(f) where the space-time plots obtained for $D=0.006$ demonstrate the front propagation starting from the same initial conditions. In the case of local coupling, $R=1$, stochastic force destroys the initial dynamics such that new domains and fronts eventually appear and disappear [Fig.~\ref{fig2}(c)]. Increasing the coupling radius allows to save the initial  fronts and domains, which didn't persist before [Fig.~\ref{fig2}~(d)] (for this reason, the curves in Fig.~\ref{fig2}~(b) are built for different noise intensity ranges, the blue curve corresponds to the longest range). Additionally, in the presence of non-local coupling we observed the collapse of domains characterized by the width comparable with the coupling radius. Thus, for larger coupling radius the exhibited front propagation finishes earlier [Fig.~\ref{fig2}~(e),(f)]. 

\section{Resonant phenomenon}
The model Eqs. (\ref{HR_ensemble}) has the following specific properties. First, the  the regime of bistability such as depicted in Fig.~\ref{fig1}(b) can be achieved in a short parameter range. Second, nullclines in Fig.~\ref{fig1}(b) are located very close to each other in the left part of the phase space. For these reasons, the localized activity bump control described above might be rather sensitive to the parameter changes. 

\begin{figure}[t!]
\centering
\includegraphics[width=0.48\textwidth]{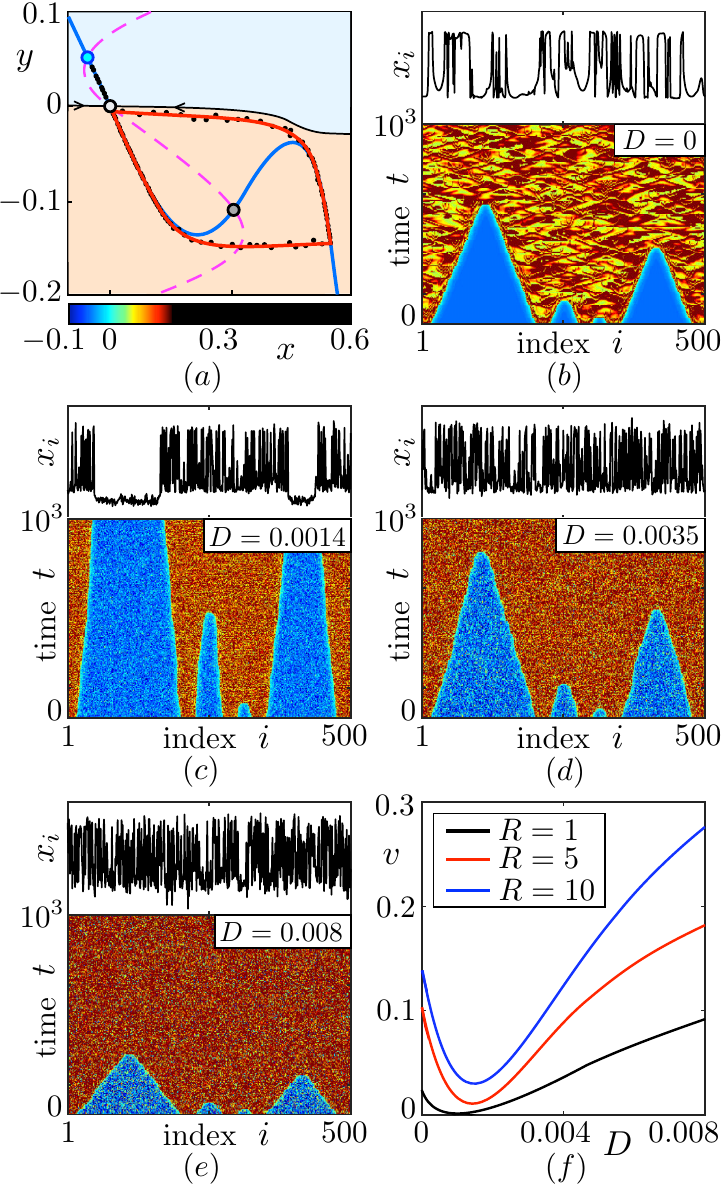} 
\caption{Dynamics in system Eqs. (\ref{SN_ensemble}): (a) Partial oscillator's phase space for $\varepsilon=0.01$, $a_1=1$, $a_3=9$, $a_5=22$, $b=2.5$, $c=20$, $d=150$. Two attractors coexist: a stable limit cycle (red closed curve) and a stable equilibrium point (blue circle). The translucent red and blue regions display the basin of attraction for the limit cycle and the equilibrium point, respectively. The light grey circle represents a saddle fixed point, black arrowed curves indicate the saddle separatrix. The dark grey circle marks an unstable steady state. Nullclines are shown by the blue solid ($dx/dt=0$) and magenta dashed ($dy/dt=0$) curves; (b)-(e) Space-time plots illustrating propagating phenomena for varying noise intensity for the fixed coupling strength $\sigma=0.7$ and coupling radius $R=10$. Other parameters are the same as in panel (a). The upper insets show the ensemble state at the last moment $t=10^3$ on an example of the variables $x_i$; (f) Dependence of the front propagation speed on the noise intensity $v(D)$ for different values of coupling radius $R$. Partial oscillator parameters are the same as for panel (a), the coupling strength is $\sigma=0.7$.}
\label{fig3}
\end{figure}  

To demonstrate the robustness of the activity bump control, as well as its different manifestation, we consider the second model: an ensemble of modified bistable spiking oscillators initially discussed in the symmetric form in Refs. \cite{semenov2016,semenov2017}, where the resonant stochastic phenomena in the chosen oscillator were demonstrated. In this work, we introduce an asymmetry into the system to achieve the coexistence of a stable limit cycle and a stable steady state. The ensemble under study admits the following form: 
\begin{equation}
\label{SN_ensemble}
\begin{array}{rl}
\varepsilon \dfrac{dx_{i}}{dt}=&-\left(1+\sqrt{2D}n_{i}(t) \right)y_{i} - a_1 x_{i}\\
&+a_3 x_i^3-a_5x_i^5,\\
\dfrac{dy_{i}}{dt}=&x_{i}+by_{i}-cy_{i}^2-dy_i^3
+\dfrac{\sigma}{2R}\sum\limits^{i+R}_{j=i-R}(y_j-y_i).
\end{array}
\end{equation}
Here $\varepsilon$, $a_{1,3,5}$, $b$, $c$, and $d$ are the partial oscillator's parameters, $R$ is a coupling radius, $\sigma$ is the coupling strength, $i=1,...,N$, $n_{i}(t) \in \mathbb{R}$ is Gaussian white noise, $D$ is the noise intensity. The partial oscillator parameters are fixed as follows $\varepsilon=0.01$, $a_1=1$, $a_3=9$, $a_5=22$, $b=2.5$, $c=20$, $d=150$, for which the partial oscillator exhibits the coexistence of a stable limit cycle (red closed curve in Fig.~\ref{fig3}~(a)) and a stable fixed point (blue circle in Fig.~\ref{fig3}(a)). There is no qualitative difference between the phase space structure of the two considered models in Fig.~\ref{fig1}(b) and Fig.~\ref{fig3}(a). The main quantitative difference  is the proximity of the attractors to the nullcline and the nullclines to each other. 

Spatio-temporal dynamics of ensemble Eqs. (\ref{SN_ensemble}) for initial conditions involving both basins of attraction consists in propagating fronts. This results in the extension of domains corresponding to self-oscillatory motions along the limit cycle (see the black points in Fig.~\ref{fig3}~(a) illustrating the instantaneous state of the ensemble) such that the spiking activity domains invade the entire available space [Fig.~\ref{fig3}~(b)]. 

Depending on the way the noise is introduced, stochastic control of the wavefront propagation manifests itself in different ways. We consider here the specific stochastic term $y_i(t)\sqrt{2D}n_{i}(t)$ (represented by multiplicative noise source, see Eqs.~(\ref{SN_ensemble})), since in this case particularly interesting features of stochastic control are revealed. 
First, the increase in noise intensity changes the front propagation speed, slowing down the spread of the spiking activity bump [Fig.~\ref{fig3}(c)]. Interestingly, in contrast to the phenomenon demonstrated in the Sec. II, further increasing the noise level in ensemble Eqs. (\ref{SN_ensemble}) speeds up the front propagation [Fig.~\ref{fig3}~(d)], which allows to obtain the localised activity bump expansion being faster than that in the deterministic system (compare Figs.~\ref{fig3}(b) and \ref{fig3}(e)). Thus, we observe a resonant effect: there is a certain noise intensity range where the front propagation velocity achieves minimum. This is illustrated in Fig.~\ref{fig3}(f), where the dependence of the wavefront propagation speed on the noise intensity $v(D)$ is shown. 

In contrast to system Eqs. (\ref{HR_ensemble}), the ensemble Eqs. (\ref{SN_ensemble})  does not exhibit noise-induced front destruction for a wide range of the noise intensity, and an increase of the coupling radius leads only to quantitative changes. Here, the growth of the coupling radius enhances the front propagation, while the resonant character of the observed effect persists [Fig.~\ref{fig3}(f)].

\section{Conclusions}
We have demonstrated that propagation phenomena in ensembles of non-locally coupled bistable spiking oscillators can be controlled by multiplicative noise. The studied noise-induced spatio-temporal dynamics is similar to the noise-controlled wavefront propagation in classical ensembles with local coupling and bistable media exhibiting the coexistence of two stable steady states.  We show that the noise-sustained control of fronts separating localized spiking activity bumps can have a monotonic character, i.e., the wavefront propagation velocity depends monotonically on the noise intensity. Interestingly, it can have a resonant character as well, i.e., there exists a certain interval of noise intensity values where the front propagation velocity reaches its minimum. 

The results obtained here show the constructive role of nonlocal coupling in preserving the propagating wavefront patterns from being destroyed by noise. This fact is an intriguing topic and will be the subject of future studies.

Certain processes in biological neuronal networks are well described by bistable models (see, for instance, bistable models for epileptic seizure description \cite{silva2003,suffczynski2004}). For this reason, the proposed approach for stochastic control of activity bump expansion can be potentially useful in the context of neurophysiology and medicine.

\section*{DATA AVAILABILITY}
The data that support the findings of this study are available from the corresponding author upon reasonable request.

\section*{Acknowledgements}
The authors sincerely acknowledge Serhiy Yanchuk for fruitful discussions and his efforts in providing valuable and helpful advice on all stages of the manucsript preparation. The authors acknowledge support by the Deutsche Forschungsgemeinschaft (DFG, German Research Foundation) -- Projektnummer -- 163436311-SFB-910. V.V.S. also acknowledges support by the Russian Science Foundation (project No. 22-72-00038). 

\nocite{*}
%

\end{document}